# PDPL Metric: Validating a Scale to Measure Personal Data Privacy Literacy Among University Students

Brady D. Lund, Nathan Brown, Ana Roeschley, Gahangir Hossain

**Abstract**

Personal data privacy literacy (PDPL) refers to a collection of digital literacy skills related to an individual's ability to understand, evaluate, and manage the collection, use, and protection of personal data in online and digital environments. This study introduces and validates a new psychometric scale (PDPL Metric) designed to measure data privacy literacy among university students, focusing on six key privacy constructs: perceived risk of data misuse, expectations of informed consent, general privacy concern, privacy management awareness, privacy-utility trade-off acceptance, and perceived importance of data security. A 24-item questionnaire was developed and administered to students at U.S.-based research universities. Principal components analysis confirmed the unidimensionality and internal consistency of each construct, and a second-order analysis supported the integration of all six into a unified PDPL construct. No differences in PDPL were found based on basic demographic variables like academic level and gender, although a difference was found based on domestic/international status. The findings of this study offer a validated framework for assessing personal data privacy literacy within the higher education context and support the integration of the core constructs into higher education programs, organizational policies, and digital literacy initiatives on university campuses.

# Introduction

Personal data privacy literacy (PDPL) can be defined as an individual's capacity to understand, critically evaluate, and manage the collection, use, and protection of their personal data in digital environments. PDPL incorporates competencies related to informed consent, privacy risk awareness, and personal data management practices. Unlike general privacy literacy, which typically focuses broadly on awareness of privacy in social, legal, or institutional contexts (Wissinger, 2017), personal data privacy literacy is focused on an individual's ability to make informed, ethical, and self-protective decisions regarding their own data within increasingly complex digital ecosystems – with emphasis on both the *personal* nature of privacy behaviors and the growing emphasis on the risks posed to private *data* in the online environment.

This paper proposes a theoretically informed scale for measuring personal data privacy literacy among university students based on six core constructs identified in the extant literature:

- perceived risk of data misuse,
- expectations of informed consent,
- general privacy concerns,
- privacy management awareness,
- privacy-utility trade-off acceptance,
- and perceived importance of data security.

This scale is measured using a 24-item questionnaire, allowing for accurate measurement while limiting survey fatigue. The proposed PDPL scale is evaluated through distribution to students enrolled at select major U.S. research universities and principal components analysis is used to validate the suitability of the constructs and individual survey items for measuring PDPL.

# Background

## Defining Personal Data Privacy Literacy

While many studies dating back decades are designed to measure privacy literacy or online privacy literacy, these literacies generally focus on privacy concerns at a broad scale or require knowledge about specific terminology or concepts rather than investigating actual privacy behaviors (Park, 2013; Sindermann et al., 2021; Weinberger et al., 2017). Personal data privacy literacy is proposed as a measure designed to evaluate privacy-related behaviors specifically linked to personal data that is exposed to the digital ecosystem. It avoids a focus on awareness of specific terminology, which may not be predictive of actual protective behaviors, and instead emphasizes the thought processes of the user as core to practical online privacy. The scale evaluated in this study relies on a foundation based on our understanding of general privacy literacy, personal data privacy, and the specific patterns of personal data privacy behavior among university students, each of which are explored in the sub-sections that follow.

### *Privacy Literacy*

Privacy literacy can be described as an individual's ability or capability to manage and control the flow of their personal information within digital and sociotechnical contexts, and to formulate their decisions within the frameworks of established privacy norms and expectations (Kumar & Byrne, 2022; Kumar, 2023). It is "a cognitive construct/state" defined as "one's level of understanding and awareness of how information is gathered and used in digital environments and how this personal information might keep or lose its private nature" (Prince et al., 2023, p. 3555). Relevant literature is in general agreement that privacy literacy encompasses both a procedural knowledge of practical strategies for managing privacy and a declarative knowledge of legal frameworks, privacy rights, and privacy regulations (Prince et al., 2023; Weinberger et al., 2017). Privacy literacy encompasses active engagement and informed decision-making and is an essential individual capability and a vital component of broader social responsibility (Kumar, 2023; Wissinger et al., 2017). Kumar and Byrne (2022) emphasize that privacy literacy extends beyond factual recall and centers on the ability to read social contexts and respond with behavior that respects privacy expectations. Kumar and Byrne (2022) further contend that privacy literacy education must be designed in collaboration and adapted to specific contexts.

Privacy literacy comprises multiple interconnected dimensions under the umbrellas of procedural knowledge and declarative knowledge, which Weinberger et al. (2017) deem "online privacy literacy (OPL)" (p. 655). Further dimensions to consider include cognitive abilities (Sindermann et al., 2021) and critical literacy (Kumar, 2023). Kumar and Byrne (2022) compare privacy literacy to learning to read as beginners start with simple descriptive tasks, but advanced learners must exercise judgment in complex, evaluative situations. The "5Ds" framework developed by Kumar and Byrne (2022) asks learners to identify data flows, explain their social context, weigh potential effects, judge consistency with norms, and choose an appropriate course of action (p. 450). Prince et al. (2023) also noted that privacy literacy requires the procedural privacy knowledge to actively apply privacy protections and the declarative knowledge to understand privacy rights legislation. However, Prince et al. (2023) took issue with the framework developed by Weinberger et al. (2017), who acknowledged that their study did not test students' knowledge of data protection laws, which left that dimension of literacy unmeasured. Prince et al. (2023) suggested "that the analysis of the impact of both declarative and procedural knowledge respectively on privacy concerns is warranted by investigating online users' awareness/knowledge and actions (privacy-protective behavior)" (p. 3554). Prince et al. (2023) defined declarative privacy knowledge as "users' awareness of websites and service providers' practices in terms of personal data collection, knowledge about technical aspects of online privacy and data protection, and knowledge about laws/legal aspects of online data protection" as well as procedural privacy knowledge such as "users' actions/skills/measures taken to alleviate their concerns over their personal information flows and protect their privacy" (p. 3554).

According to Sindermann et al. (2021), people who are more conscientious are likely to seek out additional information, which leads to higher levels of online privacy literacy (p. 1457). The authors situate privacy literacy within the "Big Five" personality traits, which include openness, conscientiousness, extraversion, agreeableness, and neuroticism (Sindermann et al., 2021, p. 1457). Sindermann et al. (2021) further reinforce the idea that individual differences in intelligence and personality shape privacy literacy outcomes, which demonstrates the influence educational and cognitive factors play in privacy competencies. Sindermann et al. (2021) argue that "crystallized intelligence," which consists of the general abilities tied to accumulated knowledge and skills, also shapes privacy literacy outcomes (p. 1456).

Kumar (2023) saw critical privacy literacy as a means to encourage collective efforts that challenge the technological and social structures that underpin surveillance, and characterizes privacy as an ongoing, socially enacted process rather than a fixed state. Kumar (2023) also asserted that privacy literacy should be seen as a "social, rather than individual, value" as access to privacy-related information notably impacts levels of privacy literacy in populations (p. 348). Meier and Krämer (2025) reported that men and individuals with higher education tend to access privacy information more easily, which enhances "literacy and self-efficacy" (p. 1183). Their findings outlined systemic disparities that can perpetuate inequalities in privacy literacy. According to Meier and Krämer (2025), privacy literacy is unevenly distributed, and older adults, women, and less-educated groups often score lower than younger or more educated peers. Meier and Krämer (2025) attribute this gap to "gender-stereotypical expectations in the labour market" (p. 1193). Advertising practices reinforce these stereotypes by marketing technology to men and health products to women (Meier & Krämer, 2025). However, Sindermann et al. (2021) found "no significant differences in the correlations between men and women" with regard to privacy literacy in their study (p. 1461). They note that knowledge of privacy issues only moderately predicts protective behavior, and that awareness does not always produce action (Sindermann et al., 2021).

Harborth and Pape (2020) indicate that privacy literacy significantly influences the adoption of privacy-enhancing technologies (PETs), and found that privacy literacy reduces generalized trust but builds confidence with specific privacy tools such as Tor. Lyu et al. (2024) explored privacy literacy in the context of privacy fatigue, and found that privacy literacy can reduce the likelihood that people will disengage from protective behaviors when their privacy is threatened. The authors define privacy fatigue as "a negative emotional factor related to the sense of losing control of personal information" (Lyu et al., 2024, p. 2007). While knowledge of privacy literacy equips users with tools to manage privacy invasions, Lyu et al. (2024) found that fatigue can still lead to reduced privacy protection behaviors. The authors suggest that people with low literacy and high fatigue may abandon privacy protections even when at risk, which helps to explain the so-called privacy paradox (Lyu et al., 2024). "People choose to disclose personal information instead of undertaking privacy protection," although "privacy literacy provides a possible way to avoid privacy fatigue" (Lyu et al., 2024, p. 2018). In this sense, privacy literacy is strengthened when individuals do not mentally succumb to the "frequent abandonment of personal information" (Lyu et al., 2024, p. 2007). "Some users actively seek solutions, adopt problem-focused coping strategies, and show less privacy protection disengagement by improving privacy literacy" (Lyu et al., 2024, p. 2019). Prince et al. (2023) also observed that increased privacy literacy results in a parallel increased concern over privacy risks, which strengthens the call for educational initiatives that couple awareness with effective strategies for mitigating risk. "Privacy literacy is a key factor determining Internet users' concerns about their personal information flows on the Web, not only in terms of protective strategies but also in terms of privacy rights knowledge" (Prince et al., 2023, p. 3554).

Privacy literacy literature reveals a complex, multidimensional construct that encompasses procedural and declarative knowledge, online privacy literacy, cognitive abilities, and critical literacy. At its core, privacy literacy empowers individuals to manage and control the flow of their personal information (Kumar & Byrne, 2022). It involves both understanding existing privacy rights and how to implement protective strategies to safeguard those rights (Prince et al., 2023; Weinberger et al., 2017). The concept of privacy literacy has also evolved beyond individual competency and can be seen as shaped by broader social and cognitive factors (Sindermann et al., 2021). Crystallized intelligence and certain personality traits contribute to individual differences in privacy literacy (Sindermann et al., 2021), while disparities in access to privacy-related information exacerbate digital inequalities (Meier & Krämer, 2025). Furthermore, critical literacy positions privacy literacy as a means for empowering individuals and communities to challenge institutional data practices and resist surveillance and datafication (Kumar, 2023).

The literature also demonstrates that privacy literacy has tangible behavioral implications. Privacy literacy influences how individuals respond to privacy fatigue and whether they remain engaged in privacy-protection behaviors (Lyu et al., 2024). It shapes their trust and willingness to adopt technological innovations that aim to protect privacy (Harborth & Pape, 2020). However, as higher privacy literacy may also heighten privacy concerns, educational efforts must foster not only awareness but also coping strategies (Prince et al., 2023). These findings underscore that privacy literacy is a vital and multifaceted component of digital life, and designing effective educational interventions, policies, and technologies requires an understanding of how privacy literacy operates at the intersection of knowledge, motivation, and social context.

*Demographic Factors in Privacy Literacy*

Research highlights differences in privacy literacy across demographic groups. Academic level is a factor as undergraduates report confidence in their abilities yet show weaker follow-through regarding protective behaviors, while postgraduates express stronger concerns yet are inconsistent in practice (Ma & Chen, 2023; Majeed et al., 2022). Residency and migration background are also notable factors. Urban students tend to perform better than rural students, which reflects the role of infrastructure and access in shaping literacy outcomes (Yi & Siqian, 2025). Students with migration backgrounds have fewer opportunities to obtain privacy-related information which reduces their sense of self-efficacy (Meier & Krämer, 2025). Gender is a somewhat inconsistent variable as women report more heightened concerns, though large-scale studies have not found reliable differences in measured literacy between men and women (Weinberger et al., 2017; Sindermann et al., 2021). However, these findings illustrate that privacy literacy is unevenly distributed across student populations.

### *Personal Data Privacy*

Personal data privacy literacy, as it relates to university students, involves both an awareness of digital privacy and the practical skills necessary for students to control and safeguard their personal data (Kardos, 2021; Lund et al., 2025). Langenderfer and Miyazaki (2009) define privacy literacy as consumers' grasp of the environments where they share data and the duties they hold in those contexts. This distinction requires both theoretical comprehension and the actionable knowledge required to manage privacy risks actively (Kardos, 2021). Unlike broader digital literacy frameworks, Kardos (2021) frames privacy literacy as an active responsibility within the information landscape, and implies a specific focus on how individuals engage with data handling and personal rights. Ma and Chen (2023) identify privacy literacy as a critical driver of protective behavior and argue for targeted educational efforts to strengthen it. A central concern in research on personal privacy literacy among university students is the "privacy paradox," a discrepancy between students' stated privacy concerns and their actual online behavior (Tsai et al., 2020, p. 8). Heinrich and Gerhart (2023) conclude that habits and trust often override awareness gains, which leaves individuals vulnerable to risky actions despite higher literacy. Ma and Chen (2023) further note that individuals frequently display "overconfidence" in their privacy knowledge, leading them to underestimate genuine online risks and, consequently, engage less frequently in effective privacy protection behavior (p. 9). When students are overconfident, privacy concerns exert less influence on protective actions, which widens the gap between perceived and real ability and exposes a critical gap between perceived understanding and actual capability (Ma & Chen, 2023).

Existing literature indicates notable differences in data privacy attitudes between undergraduate and postgraduate students. Undergraduate students typically balance privacy concerns against convenience by acknowledging potential risks yet still accepting compromises when engaging with online platforms (Ray & Feinberg, 2021). Ray and Feinberg (2021) documented this perspective in noting that an undergraduate explicitly expressed willingness to sacrifice convenience for privacy, which highlights the importance placed on safeguarding personal information. Conversely, postgraduate students frequently display heightened sensitivity to privacy invasions, especially in cases where educational platforms intersect with personal data or external parties (Tsai et al., 2020). Responses by postgraduates in the research by Tsai et al. (2020) reflect stronger demands for consent, autonomy, and transparency in institutional data handling practices. Despite this greater sensitivity, postgraduate students still face significant challenges in consistently applying their privacy knowledge (Kardos, 2021). Kardos (2021) found that postgraduate law students frequently overlooked accidental disclosures and interpreted personal data breaches narrowly as acts of criminal intent. Specifically, Kardos (2021) reported that only one student interviewed recognized that data breaches could occur accidentally, without malicious intent, while other respondents solely linked breaches to unlawful acts. Kardos (2021) further noted that most postgraduate students frequently overlook accidental breaches and lack sufficient knowledge to manage them effectively.

According to Tsai et al. (2020), students place implicit trust in universities to manage their personal data responsibly, and often assume institutional goodwill rather than actively engaging with how education analytics systems collect, process, or share their information. This implicit trust, combined with a lack of engagement with data policies, contributes to a knowledge gap regarding how personal data is collected, processed, and shared within the institution (Tsai et al., 2020). Chen and Chen (2025) similarly found that while students express concern about privacy in online education systems, many lack clarity about how their data is handled, which suggests a need for improved privacy education strategies and transparency. According to Ma and Chen (2023), this reflects a broader cognitive gap as students frequently overestimate their privacy knowledge compared to objectively measured literacy levels, which significantly reduces practical effectiveness and weakens data protection. According to Ma and Chen (2023), students often overestimate their privacy control skills compared to their actual literacy levels. Ma and Chen (2023) advocate for targeted education and training to improve student privacy literacy, while Ray and Feinberg (2021) emphasize the importance of integrating privacy into information literacy education instruction to foster meaningful student engagement.

Majeed et al. (2022) found that postgraduate students expressed significant privacy concerns in online learning environments, but these concerns did not consistently translate into privacy protection behaviors, highlighting the need for educational interventions. Chen and Chen (2025) further observed that student trust in online learning environments is shaped by their perceptions of how securely their data is managed and whether privacy risks are adequately communicated. Koohang et al. (2018) also identified persistent student concerns about unauthorized access and secondary use of personal data on social media, yet these concerns did not consistently lead to protective behaviors and reflected the type of inconsistency often described in privacy paradox literature. To effectively address these discrepancies, Prinsloo et al. (2022) argue that effective student data protection requires transparency, active student involvement, and systemic approaches rather than technology alone to support meaningful privacy protection in educational contexts.

Educational strategies and curriculum development could effectively bridge the gap between privacy awareness and behavioral practice. Ray and Feinberg (2021) support the integration of privacy literacy into existing curricula, as privacy literacy influences both academic work and personal information sharing. Blackmon and Major (2023) call for institutions to maintain clear, transparent policies on data collection and involve students in shaping privacy rules. While not framed in terms of privacy literacy, these strategies promote foundational conditions by ensuring students are informed, engaged, and capable of understanding and managing their digital privacy in educational settings through institutional transparency, student agency, and informed participation, which are key conditions for integrating privacy awareness (Blackmon & Major, 2023). Hartman-Caverly et al. (2023) recommend student-centered instruction that cultivates awareness, self-reflection, critical thinking, and judgment in digital privacy contexts, and underscore the significance of inclusive, responsive approaches in privacy literacy education that respect student autonomy and personal experience. Qualitative student insights provide valuable context, as illustrated by Ray and Feinberg (2021), and highlight the intentional trade-offs students often make between maintaining an online presence and safeguarding their private, personal information. Tsai et al. (2020) highlight the need for universities to be transparent about why they collect data, who can access it, and how anonymity is preserved rather than assume students inherently understand such systems.

*Demographic Factors in Personal Data Privacy*

Literature on personal data privacy literacy (PDPL) shows demographic variations that are similar to general privacy literacy, although with different emphases. Academic level remains a central variable as undergraduates accept trade-offs and value convenience even when it compromises privacy (Ray & Feinberg, 2021). Postgraduate students display gaps such as overlooking the possibility of accidental breaches (Kardos, 2021). Despite higher levels of concern, postgraduates further fall short in consistently applying protective strategies (Majeed et al., 2022). Residency and migration backgrounds also shape how students engage with PDPL. Rural students demonstrate weaker familiarity with security tools and practices compared to urban peers (Yi & Siqian, 2025). Students from migration backgrounds encounter additional barriers to privacy information, which undermines their ability to act with confidence (Meier & Krämer, 2025). Gender patterns echo those in general privacy literacy as women are more cautious in disclosure, yet overall literacy levels show no systematic differences (Weinberger et al., 2017; Sindermann et al., 2021). Demographic variations, then, influence not only how students perceive risks but also how effectively they respond to them.

**Examining the Core Constructs of Personal Data Privacy Literacy**

This study theorizes that personal data privacy literacy can be described as a single competency that is comprised of six major constructs that relate to privacy perceptions and actions on the part of individuals: their perceived risk of data misuse, their expectations of informed consent, their general privacy concerns, their privacy management awareness, their privacy-utility trade-off acceptance, and their perceived importance of data security. The following sub-sections explore each of these constructs, their basis in the literature, and how they may contribute to our understanding of personal data privacy literacy.

*Perceived Risk of Data Misuse*

Perceived risk of data misuse refers to an individual's recognition of the potential negative consequences that may emerge from a party's unauthorized or unethical use of that individual's personal information. This construct helps to assess the extent to which individuals are able to critically evaluate the implications of sharing their personal information and take appropriate actions, which has been noted by authors such as Hysa et al. (2023) as an important aspect of privacy literacy. Given the historical frequency of data misuse by companies and government entities, some caution or skepticism when it comes to sharing data with these groups is healthy and a strong indication that the individual takes their personal privacy seriously (Beldad et al., 2011; Nguyen & Simkin, 2017).

*Expectations of Informed Consent*

Expectations of informed consent capture the extent to which individuals believe they should be made aware of, and agree to, how their personal data is collected, processed, and used. This construct reflects a key component of personal data privacy literacy: understanding one's rights and advocating for transparency and ethical standards in data practices. An understanding of what one should be questioning when it comes to privacy can be said to be a major factor that distinguishes the literate from the illiterate (Hagendorff, 2018; Trepte et al., 2015). Those who lack this ability are more likely to take a noncritical approach to adopting new digital tools, which places them and their data at a tremendous risk (Lund et al., 2023).

*General Privacy Concern*

General privacy concern refers to an individual's overall sense of unease or apprehension about the widespread collection and use of personal information by organizations. This construct reflects awareness of potential privacy violations and motivates protective behaviors, representing an emotional but foundational aspect of data privacy literacy. Having some general concern about one's privacy being breached is a natural and generally positive inclination in promoting protective behaviors (Baruh et al., 2017). Concerns about privacy are often directly reflected in actions, such as use-intention related to technologies that pose a privacy risk (Adhikari & Panda, 2018; Harborth & Pape, 2020).

*Privacy Management Awareness*

Privacy management awareness involves the knowledge and behaviors individuals use to protect their personal information, such as adjusting privacy settings, reading data policies, and staying informed about digital security practices. This construct demonstrates how well individuals can actively manage their data privacy, a critical skill in maintaining personal data security. Historically, the emergence of social media posed a major threat to personal online privacy, with many individuals failing to realize how the information they share online may be used by developers, fellow users, and third parties. Over time, though, research has shown that privacy management skills have improved, especially among traditional "college age" young adults (Chen & Chen, 2015; Madden, 2012). However, the emergence of generative artificial intelligence tools like ChatGPT present new challenges for privacy management, as many users fail to realize there are settings they can adjust in order to avoid data collection and use of their prompts in training future iterations of the models (Alawida et al., 2024; Wu et al., 2024).

*Privacy-Utility Trade-off Acceptance*

Privacy-utility trade-off acceptance describes an individual's willingness to exchange personal data for benefits such as access to services or personalized experiences. This construct illustrates a person's capacity to evaluate and make informed decisions about the risks and rewards of data sharing, which is critical to contextualizing data privacy literacy. The privacy-utility trade-off has been examined extensively in system privacy research (Li & Li, 2009; Sreekumar & Gündüz, 2019; Valdez & Ziefle, 2019). These studies suggest a high level of individual variation in privacy-utility trade-off related to risk tolerance and understanding of a system's value.

*Perceived Importance of Data Security*

The perceived importance of data security refers to how much value individuals place on the use of tools and practices that protect personal information from unauthorized access or breaches. This construct reflects the technical understanding and prioritization of secure data handling, which is essential for informed and responsible privacy behavior. Perceived importance of data security is another well-understood construct that has been found to relate to threat avoidance behaviors and adoption of emerging technologies (Huang et al., 2010; Suh & Han, 2003). Some studies have found perceived importance of security to be context-dependent, which informs this study's focus on data privacy and security within a higher education context (Shah et al., 2014).

**Research Problem, Questions, and Hypotheses**

As university students face an era of big data and generative artificial intelligence where personal data is regularly collected, stored, and used by a wide range of organizations for unclear purposes, individuals' ability to understand, manage, and make their own informed decisions about their data privacy is vital. Though awareness of data privacy concerns has grown in recent years, there remains a limited understanding of how well-equipped students are in terms of their personal data privacy literacy. Current measurement tools for privacy literacy often fail to capture its multifaceted nature, leading to a need for a reliable and valid instrument that reflects the complex competencies involved in personal data privacy decision-making. The purpose of this study is to 1) refine and validate a concise and multidimensional scale for measuring personal data privacy literacy, and 2) examine the relationship of this measure to a few demographic factors relevant to the university student.

**Research Questions**

1. To what extent do the six identified dimensions (perceived risk of data misuse, expectations of informed consent, general privacy concern, privacy management awareness, privacy-utility trade-off acceptance, perceived importance of data security) represent a unified construct of personal data privacy literacy?
2. Does personal data privacy literacy differ significantly based on demographic variables such as academic level, gender, and residency status?

## Methods

Following the identification of the relevant constructs for the personal data privacy literacy scale, a set of questions were developed to measure each construct. Several questions in this survey were adapted from Addae et al.'s (2017) Adaptive Cybersecurity Scale and Trepte et al.'s (2014) Online Privacy Literacy Scale, with revisions made to better assess the theorized constructs informing personal data privacy literacy. Wording of the questions was reviewed by the research team and cybersecurity graduate students to ensure clarity. Each question utilized a five-point Likert scale, with the options ranging either from "strongly disagree" to "strongly agree" or "never" to "always." For each of the six constructs, only four questions were selected, allowing for an efficient measurement of the central phenomena with a scale consisting of just 24 items. Additional demographic questions – covering educational attainment, gender, and residency status – were included to help ensure that the scale's results were not disproportionately influenced by any single demographic group.

Data collection for this study occurred in Spring 2025. An electronic version (Qualtrics) of the survey instrument was distributed via direct mail to students at several large research universities in the United States. Participants were encouraged to share the survey link with colleagues. A paper version of the survey was also made available to interested participants. All participants were currently enrolled university students. 369 complete responses were received. These responses were exported to SPSS for statistical analysis.

To assess the structure and validity of the personal data privacy literacy scale, a series of statistical analyses were conducted. Principal components analyses (PCA) were performed on each of the six identified constructs to confirm their unidimensionality. Internal consistency reliability for each construct was evaluated using Cronbach's alpha, with all constructs demonstrating acceptable to high reliability (α values > .70). A second PCA was then conducted using the composite scores of the six constructs to explore whether they represent a broader, unified construct of personal data privacy literacy. Finally, a series of independent samples t-tests were conducted to examine whether personal data privacy literacy scores differed based on the demographic factors of academic level (undergraduate vs. graduate), gender, or residency status (domestic vs. international).

# Results

**Principal Components Analysis of Personal Data Privacy Constructs**

A principal components analysis (PCA) was conducted to examine the underlying structure of six constructs related to personal data privacy literacy. Table 1 presents the survey items, standardized factor loadings, and eigenvalues for each construct. All items were subjected to individual PCAs to verify unidimensionality. In each case, a single factor emerged with an eigenvalue greater than 1, supporting the internal consistency and cohesion of each construct.

**Table 1. Personal Data Privacy Literacy Constructs, Loadings, and Survey Questions**

| Construct | Questions | Loading |
|---|---|---|
| Perceived Risk of Data Misuse (Eigenvalue = 2.38) | 1. I believe personal information can be misused to harm individuals. | .729 |
| | 2. I believe external organizations place significant value on my personal information. | .645 |
| | 3. I am concerned that databases storing my personal data may be vulnerable to unauthorized access. | .868 |
| | 4. I worry that inaccurate information may be linked to my identity due to data breaches. | .823 |
| Expectations of Informed Consent (Eigenvalue = 2.59) | 1. Organizations should clearly disclose how they collect, process, and use personal data. | .744 |
| | 2. I am bothered when organizations do not explain how my personal information will be handled and secured. | .833 |
| | 3. Data handlers have a responsibility to ensure that personal information is used ethically and appropriately. | .848 |
| | 4. Researchers should obtain my consent before accessing my personal data. | .787 |
| General Privacy Concern (Eigenvalue = 2.88) | 1. I am concerned about the excessive collection of my personal information by various organizations. | .845 |
| | 2. I worry that the personal information I provide may be exposed or leaked. | .910 |
| | | .876 |
| | 3. I am hesitant to share personal data because it might be used in ways I did not anticipate.<br>4. I am bothered when I lack control over how my personal data is collected, used, and shared. | .754 |
| Privacy Management Awareness (Eigenvalue = 2.29) | 1. I adjust privacy settings when creating online profiles.<br>2. I review the privacy policies of organizations before sharing personal information.<br>3. I actively stay informed about new personal data protection policies.<br>4. I take steps to protect my personal data from unauthorized access. | .756<br>.714<br>.764<br>.792 |
| Privacy-Utility Trade-off Acceptance (Eigenvalue = 2.44) | 1. My need to access services often outweighs my personal privacy concerns.<br>2. I am willing to share personal information to support government policy and decision-making.<br>3. I value the personalized services I receive in exchange for sharing my data.<br>4. I would be open to paying for services that ensure the privacy of my personal data. | .708<br>.824<br>.879<br>.695 |
| Perceived Importance of Data Security (Eigenvalue = 1.87) | 1. I use security tools to protect my personal data.<br>2. I am concerned about online data being linked to my publicly available offline information.<br>3. I believe the risks of sharing personal data outweigh the potential benefits.<br>4. I believe stronger security measures are necessary to ensure the protection of my personal data. | .760<br>.816<br>.654<br>.545 |

**Unified Factor Structure for Personal Data Privacy Literacy**

A second PCA was conducted using the six construct scores to determine whether they reflect a broader underlying dimension of personal data privacy literacy. Sampling adequacy was verified using the Kaiser-Meyer-Olkin (KMO) test, which yielded a value of .741, indicating a "middling" yet acceptable fit. Bartlett's Test of Sphericity was also significant, $\chi^2(15) = 421.373$, $p < .001$, confirming that the correlation matrix was appropriate for factor analysis.

PCA revealed a single component with an eigenvalue of 3.08, explaining 51.33% of the total variance. All six variables loaded positively on this component, with loadings ranging from .525 to .882 (Table 2). Based on the Kaiser criterion (eigenvalue > 1) and visual inspection of the scree plot, only one component was retained. These results support the interpretation of the six variables as indicators of a unified personal data privacy literacy construct.

**Table 2. Factor Loadings for Personal Data Privacy Literacy Construct**

| Construct | Loading |
|---|---|
| Perceived Importance of Data Security | .882 |
| Perceived Risk of Data Misuse | .749 |
| Expectations of Information Consent | .755 |
| General Privacy Concern | .783 |
| Privacy Management Awareness | .525 |
| Privacy-Utility Trade-Off Acceptance | .531 |

*Relationship Between Demographics and Personal Data Privacy Literacy*

A series of independent samples t-tests were conducted (Table 3). These tests found no statistically significant difference in personal data privacy based on academic level (undergraduate vs. graduate) or gender. However, residency status (domestic vs. international student did show a significant difference, with domestic students having PDPL scores that were significantly higher than those of international students who participated in this study.

**Table 3. Test Statistics for Personal Data Privacy Literacy and Participant Demographics**

| Variable | Test Statistic | P-Value |
|---|---|---|
| Academic Level | 1.542 | .215 |
| Gender | 1.489 | .223 |
| Residency Status | 6.297 | .013 |

## Discussion

The purpose of this study was twofold. The first aim was to determine whether the six theoretically identified constructs of perceived risk of data misuse, expectations of informed consent, general privacy concerns, privacy management awareness, privacy-utility trade-off acceptance, and perceived importance of data security could be considered unified indicators of a construct of personal data privacy literacy (PDPL). Second, the study sought to examine whether any significant differences in PDPL among university students could be tracked to their demographic characteristics.

**Addressing Research Question 1**

The findings of the principle components analyses provide a strong positive indication for research question 1 and hypotheses 1 and 2 for this study. For all six of the theoretically identified constructs, high loadings were identified for the four questions, with eigenvalues significantly exceeding a 1.0 threshold. In most cases, the loadings for each individual question exceeded a .700 threshold. The weakest eigenvalue and loadings were associated with the "perceived importance of data security" construct; however, these still meet an acceptable level of support (Hair et al., 2018).

The second PCA conducted on the composite scores from the six constructs found a single dominant construct with a significant eigenvalue of 3.18. This finding suggests that the six identified constructs are meaningfully interrelated and collectively represent the higher-order construct of PDPL. The factor loadings range, ranging from .479 to .899, and the high level of internal consistency indicated by the Cronbach's alpha values, show strong empirical support for treating PDPL as a unified, multidimensional construct, which can be suitably measured using the 24-item scale survey. This contribution is significant for advancing research relating to personal data privacy literacy, which has been identified as a core literacy for navigating the modern digital infrastructure but is often not directly measured in survey-based studies (Hillman, 2022; Suddekunte et al., 2024).

These findings provide promising paths of inquiry for future research regarding the "privacy paradox," or discrepancy between university students' stated privacy concerns and their actual online behavior (Tsai et al., 2020, p. 8). These findings do not dispute that user "overconfidence" in their privacy literacy can lead to risky behaviors (Ma & Chen, 2023, p.9). However, unified indicators of a construct of personal data privacy literacy could bring further insights into situations where "individuals still engage in risky behaviors" despite an increase of personal privacy education (Heinrich & Gerhart, 2023, p. 49). With the "privacy-utility trade-off acceptance" construct as part of holistic PDPL, online behaviors perceived as risky need to be contextualized within a larger PDPL context.

**Addressing Research Question 2**

The findings of the independent samples t-tests show no statistically significant differences in PDPL across the demographic categories of academic level and gender. However, a significant finding was found for residency status. This finding aligns with some existing literature that points to differences among domestic and international students for reasons such as linguistic differences and different conceptions of privacy and privacy behaviors when compared to the U.S.-based context in which this scale was primarily developed (Jover, 2018; Sin & Kim, 2013). This may suggest that the scale is less useful for measuring the personal data privacy literacy outside of the U.S. context/definition of this concept.

Future research may explore PDPL using this scale in relation to other demographic attributes of university students, such as age or the type of institution in which the student is enrolled, which have been shown in some studies to be more predictive of privacy and data literacy behaviors (Kim et al., 2025; Mutimukwe et al., 2022). While Sindermann et al. (2021) found "no significant differences in the correlations between men and women," other studies have found gender-based discrepancies in privacy literacy (p. 1461). Surveying other demographic groups can support the development of applied research projects to make privacy literacy more attainable across demographic markers as "multiple studies imply that older persons, less educated persons, and women have a lower privacy literacy than younger individuals, more educated ones, and men" (Meier and Krämer, 2025, p. 1184). Furthermore, in the domain of critical privacy literacy, unified and interrelated indicators are compatible with conceptualizations of privacy as a "dynamic practice" (Kumar, 2023, p. 348).

**Implications for Research and Practice**

The findings of this study offer several implications for academic research as well as practical applications in the areas of digital literacy, data ethics, and privacy education. The results provide empirical support for the notion of personal data privacy literacy (PDPL) as a unified yet multidimensional construct. The validated 24-item measure supports consistent and replicable future research relating to the different dimensions of PDPL, demographics of users, and its influence on online behaviors and responses to data tasks. Its multidimensional attributes make the scale applicable to conceptualizations of privacy literacy as "the skills to interpret social situations and act in ways that align with privacy" (Kumar & Byrne, 2022, p. 446). As such, this measure can be integrated into broader evaluations of digital literacy skills and behaviors, particularly among university students (the context of the present study).

The findings are congruent with Kumar and Byrne's (2022) approaches to privacy literacy education as a "collaborative, contextually driven process" (p. 446). In terms of practical applications of this research, the PDPL measure scale can be used to assess baseline student awareness of data privacy concepts and inform privacy curriculum development. Prince et al. (2023) argue that privacy curricula need to result in both awareness and strategies for best practices, while Lyu et al. (2024) explain that effective privacy literacy can empower privacy-protection behaviors. The six components of the PDPL construct — importance of data security, risk of data misuse, informed consent, general privacy, privacy management, and privacy-utility trade-off — provide a good structure for a data privacy literacy instructional program. Universities may integrate PDPL into institutional assessments like first-year student surveys in order to track the evolution of students' understanding of data privacy and help inform institutional policies relating to data ethics and responsible technology use.

This PDPL Metric improves upon prior models, which have more limited definitions of this literacy, or include only certain aspects of personal data privacy literacy within a measurement of broader phenomenon. This metric effectively measures the singular concept of personal data privacy literacy across 24 questions, allowing the instrument to be deployed efficiently to a large population as either a stand-alone survey or one component of a larger survey measuring several phenomena related to the privacy behavior or digital lives of university students.

**Limitations**

There are several limitations to note for this study. The sample used in the study was limited to students from a few U.S.-based research universities, which may not represent student populations at other universities or other populations outside of the higher education context. The relatively small sample size limits the statistical power and may have created limitations to measuring demographic differences in responses. Additionally, the structure of the PDPL scale may introduce the possibility of social desirability bias influencing responses. Future development on the scale might involve combining the Likert items with problem-based questions to evaluate certain aspects of personal data privacy literacy.

## Conclusion

Through an empirical validation of the personal data privacy literacy construct and development of the PDPL measurement scale, this research provides a framework for advancing our understanding of how individuals — particularly students — navigate personal data privacy in complex digital environments. It advances past existing models by providing a short and adaptable scale by which to evaluate this phenomenon among university students. By presenting this multidimensional measure of privacy literacy, this study also creates opportunities for expanding our evaluation of the effectiveness of interventions aimed at enhancing individuals' capacity to make informed data-related decisions. This PDPL construct further offers a path for the integration of privacy literacy into higher education programming, organizational policy, and digital literacy initiatives.